**Localised soft modes and the supercooled liquid's irreversible passage through its configuration space**

Asaph Widmer-Cooper[1,2], Heidi Perry[3], Peter Harrowell[2] and David R.Reichman[3]

[1] *Materials Science Division, Lawrence Berkeley National Laboratory, Berkeley, California 94720, USA*
[2]*School of Chemistry, University of Sydney, Sydney, New South Wales 2006, Australia*
[3]*Department of Chemistry, Columbia University, 3000 Broadway, NY, NY, 10027, USA*

Abstract

Using computer simulations, we show that the localized low frequency normal modes of a configuration in a supercooled liquid are strongly correlated with the irreversible structural reorganization of the particles within that configuration. Establishing this correlation constitutes the identification of the aspect of a configuration that determines the heterogeneity of the subsequent motion. We demonstrate that the spatial distribution of the summation over the soft local modes can persist in spite of particle reorganization that produces significant changes in individual modes. Along with spatial localization, the persistent influence of soft modes in particle relaxation results in anisotropy in the displacements of mobile particles over the timescale referred to as β relaxation.

# 1. Introduction

In his 1969 essay on viscous liquids and the glass transition, Goldstein [1] described the supercooled liquid as existing, predominantly, around local potential minima. The slow relaxation kinetics in such states corresponds, in Goldstein's picture, to the quasi-localized processes by which the transition between local minima is accomplished. This picture remains, to this day, both undisputed and unfulfilled. While a considerable body of computer simulations have verified the importance of local potential minima in describing the thermodynamics of supercooled liquids [2], the real space description of the transitions between these minima remains an open question. It is difficult to imagine a satisfactory account of the relaxation processes of these low temperature liquids without a description of the real space, physical processes by which they occur.

Collective dynamics in supercooled liquids are typically characterised by dynamic heterogeneities [3]. The latter term refers to the observation of long-lived domains of particles distinguished by different relaxation kinetics. Where the configuration space representation favoured by Goldstein provides a formally elegant and complete description of the collective dynamics of a supercooled liquid, the study of dynamic heterogeneities presents the real space details of the cooperative dynamics with a degree of complexity that borders on the zoological. Simply describing these heterogeneities represents a substantial intellectual challenge. This description allows us to develop an account of the physical process responsible for slow structural relaxation in the form of explicit questions involving dynamic heterogeneities. In this paper we address two such questions. 1) What motions matter with respect to relaxation? How do we separate out those motions directly involved with the irreversible relaxation we are interested in? 2) What aspects of the structure correlate with the heterogeneity of the motions associated with relaxation? As we shall discuss, some feature of structure in the particle configuration must be responsible for this distribution of relaxation rates. In this paper we provide evidence that irreversible relaxation of configurational correlations takes place through the action of quasi-localised soft modes. A preliminary report of this work has appeared recently [4].

## 2. Background

## 2.1 The Iso-Configurational Ensemble and Short Time Heterogeneities

A fruitful computational approach to identifying the structural origin of dynamical heterogeneity has been put forward in the dual notions of an "iso-configurational ensemble" and "propensity for motion" [5]. The iso-configurational ensemble refers to the ensemble of trajectories that are run from an identical configuration of particles with random initial momenta sampled from the equilibrium Boltzmann distribution. Propensity refers to the mean squared displacement of individual particles when averaged over the ensemble at a given time scale. The heterogeneous character and increased clustering exhibited in the spatial propensity maps established that the spatial distribution of these dynamic heterogeneities can be explicitly attributed to structural features, as yet unidentified.

Using the iso-configurational ensemble, we can establish the component of the spatial distribution of the dynamics that is determined by the initial configuration. It is this component of the dynamics whose cause must lie in the particle arrangements of the initial configuration. The search for this structural explanation, however, has met with its share of failures. In the 2D soft disk mixture, the spatial distribution of various local properties of the structure: topology, composition, free volume and single particle energies in the local potential minimum, do not show sufficient correlation with the spatial distribution of the propensity to warrant attaching any causal connection [6]. The conclusion drawn from this lack of success was that the collective character of relaxation required a similarly collective structural feature as its cause. In lieu of a clear idea of what nonlocal structural measure to use, it was shown that maps of the local Debye-Waller factor (the iso-configurational mean squared displacement over a short time interval, $10\tau$, corresponding to the early $\beta$ region) provided an excellent prediction of where in space the slower relaxation events would take place [7]. The particles, in contrast to ourselves, are 'capable' of quickly establishing which structural domains are likely to eventually experience large particle displacements. While some of these short time displacements exceeded the range associated with the harmonic approximation, these results suggested that the normal modes of the local potential energy minima associated with an initial configuration could provide the sought after connection between dynamic heterogeneity and structure. Before reviewing the literature on normal modes and the glass transition, we shall, in the next Section, consider the issue of irreversibility and local dynamics in a glass forming liquid.

## 2.2 Irreversibility

Relaxation in a many body system arises when fluctuations occur that do not revert to their original state within a time short enough to ensure coherence of the initial state. Such fluctuations can be called 'irreversible'. In this context, we shall refer to a motion as being 'irreversible' if the probability of observing a reversal lies below some threshold value. Irreversibility, if the concept can be sensibly attributed to small groups of particles, can provide a generic identifier of motions that contribute to relaxation and, therefore, is an attractive means of sorting relevant from irrelevant types of motion.

Doliwa and Heuer [8] identified metasbasins in configuration space purely in terms of the irreversibility of the collective behaviour of small (60 particle) systems. Specifically, they determined the probability of return to previously visited local potential minima when a trajectory was interrupted and restarted with random momenta. This brings us to the question of the choice of the number of particles required to assess irreversibility. A large system will contain many dynamically independent regions so that irreversibility is unavoidable but uninformative with respect to the correlated dynamics. The 60 particles used in ref. [8] were selected as the smallest number of particles for which finite system size effects remained negligible. If, instead of employing a small system, we defined sub-domains in a large system, finite size effects are not relevant. With the disappearance of this lower bound the smallest sample size can be reduced to a particle and its nearest neighbours. In Section 4.1, we define a single particle measure of irreversibility. Previously, Vollmayr-Lee [9] resolved individual particle motions in a supercooled liquid mixture into reversible and irreversible jumps and demonstrated that the ratio of the latter to the former decreased on cooling.

## 2.3 Instantaneous and Quenched Normal Modes in Supercooled Liquids

There is a considerable literature on the application of normal modes to supercooled liquids and amorphous solids. Much of the attention has focussed on the density of states and dispersion curves. In this Section we provide a brief review of that component of the literature which has addressed the link between soft modes, their spatial distribution and large amplitude particle displacements.

In 1991 Schober and Laird [10,11] demonstrated the existence of low frequency quasi-localised modes (QLM) in model glasses. The qualification "quasi-" refers to the fact that these modes typically include an extended component. Ref. [12] provides an excellent discussion of this point. On heating, these workers found [13,14] that the relaxation events have similar participation ratios to those found in the QLM's, and a correlation coefficient, obtained from considering the expansion of the relaxation displacements in terms of the modes, was found to have values around 0.5 (where $1/N$ means no correlation and 1 means perfect correlation). Note that the comparison in refs. [13, 14] was between modes at $T = 0$ and relaxation events at a non-zero temperature. Recently, Brito and Wyart [15, 16] have shown that for small systems of hard disks there is a considerably better overlap between the relaxation displacement and even a single QLM if effective free-energy normal modes are identified immediately prior to the relaxation event. The authors of this work [15, 16] point out that these results establish that the collective displacements in their small systems are a consequence of essentially a single soft mode as opposed to being the accumulation of a sequence of separate sets of displacements.

The description of the spatial character of the normal modes of amorphous states remains a challenge. In a 2D spring lattice with disorder in the force constants, the spatial distribution of the prelocalized modes exhibits multi-fractal scaling [17]. Vollmayr-Lee and Zippelius [18] have examined the size distribution of clusters made up of particles exhibiting large amplitude vibrational motion. An alternate perspective on the description of the spatial character of particle motions in glasses comes from the study of motion under an applied strain. Barrat and coworkers [19,20] examined aspects of the spatial structure of these modes in the context of studying the non-affine character of local strain in glasses. More recently, these workers [21] have argued that the decomposition of the strain fields into a coarse grained component and the fluctuations about this mean provides a more clearly resolved picture of the localization of the soft modes. Such spatial structures attract the same question regarding cause that we have already discussed in the context of dynamic heterogeneities. Luchnikov et al [22] have presented an analysis in support of the proposition that the QLM are associated with local geometric instabilities, measurable in terms of the degree of tetrahedrality. In perhaps the most explicit statement to date on the connection between structure and vibrations in a glass, Guerdane and Teichler [23] have suggested that the Boson peak (an enhancement of the vibrational density of states at low frequency) is a consequence of the transitions between a few local coordination polyhedra.

### 3. Model and Analysis

For our model of a glass-forming liquid, we use a two-dimensional (2D) equimolar binary mixture of particles interacting via purely repulsive potentials of the form

$$u_{ab}(r) = \varepsilon \left[ \frac{\sigma_{ab}}{r} \right]^{12} \qquad\qquad (1)$$

where $\sigma_{12} = 1.2$ x $\sigma_{11}$ and $\sigma_{22} = 1.4$ x $\sigma_{11}$ [24]. All units quoted will be reduced so that $\sigma_{11} = \varepsilon = m = 1.0$ where m is the mass of both types of particle. Specifically, the reduced unit of time is given by $\tau = \sigma_{11} (m/\varepsilon)^{1/2}$. The average collision time at $T = 0.4$ in the binary mixture is $0.1\tau$. The reduced unit of temperature is $k_B/\varepsilon$. A total of $N = 1024$ particles were enclosed in a square box with periodic boundary conditions. Molecular dynamics simulations were performed in the NPT ensemble using a Nosé-Poincaré-Andersen algorithm developed by Sturgeon and Laird [25]. The structural or 'alpha' relaxation time $\tau_\alpha$ is defined as the time required for the self intermediate scattering function $F_s(q,t)$,

$$F_s(q,t) = \frac{1}{N} \left\langle \sum_{j=1}^{N} \exp\{ i\vec{q} \cdot [\vec{r}_j(0) - \vec{r}_j(t)] \} \right\rangle \qquad (2)$$

to decay to a value of $1/e$. The magnitude of the wavevector $q$ is set equal to the value at the first Bragg peak.

For the normal mode analysis, the inherent structure of each configuration was found using the conjugate gradient method. The dynamical matrix of the inherent structure was then defined as $\mathbf{D} = \dfrac{\partial^2 \Phi(r)}{\partial r_i^k \partial r_j^k}$ where $r_i^k$ is the $k^{th}$ component of the position $\mathbf{r}_i$ of particle $i$, $\Phi(r) = \sum_{i=0}^{N} \sum_{j=1, j \neq i}^{N} u_{ab}(r_{ij})$ , $u_{ab}$(r) is the intermolecular potential from Eq.1 and $r_{ij}$ = |$\mathbf{r}_i$-$\mathbf{r}_j$|. The dynamical matrix was diagonalized using the template numerical toolkit [26].

To visualize the spatial distribution of the propensity, it is useful to remove the additional complexity of the configuration and use contour plots. As the data points are located at irregularly spaced particle coordinates, it is necessary to interpolate between them. We have used the modified version of Shepard's method as implemented in the NAG libraries.

## 4. Results

### 4.1 The Spatial Distribution of Irreversible Relaxation

To differentiate reversible vibrations from motion corresponding to an irreversible change in particle arrangements, we shall consider the changes in nearest neighbour topology. First, we establish how many nearest neighbours must be lost around a given particle before the probability of the tagged particle recovering its original environment falls below 5%. The probability that a particle could lose $x$ of its original neighbours and subsequently *fail* to recover $n$ of those lost neighbours was determined as follows. A hundred runs were carried out from a given starting configuration with momenta assigned randomly from an equilibrium Boltzmann distribution at the appropriate temperature. Run intervals of 1000τ and 2000τ were used. The initial neighbours of each particle were determined by using a cut-off distance equal to the distance to the first minimum in the species-appropriate radial distribution function.

During the course of each trajectory, the neighbours of each particle were monitored. When particle $i$ first lost $x$ of its initial neighbours (with $x$ going from 1 to 5), the minimum number $n$ of *lost neighbours* that *failed* to be recovered in the remainder of the trajectory of particle $i$ was recorded. Complete recovery of the original neighbourhood corresponds to $n = 0$ lost neighbours that fail to be recovered. Note that a particle would be 'marked' as having $n = 0$, for example, even if it only recovered those neighbours for a very short time interval during the subsequent trajectory. Collecting data from each particle for all 100 runs, we plot the fraction of total observations corresponding to each value of $n$ in Fig. 1.

Losing a lot of bonds (i.e. large $x$) would be expected to take a longer time than losing a few bonds. That would mean that there would be less time left in the trajectory for the large $x$ particles to recover their original neighbours. To see how serious this effect is, we repeated the calculations for a time interval of $2000\tau$, twice that used in Fig. 1. We found that the distributions shown in Fig. 1 were shifted to the right (i.e. towards higher $n$). We infer that increasing the run time will not alter our conclusion that the loss of 4 initial neighbours constitutes irreversible reorganization.

The choice of time interval ($200\tau$, in this case) over which to observe particle loss introduces a coarse graining of time. One might ask how the time intervals over which particles first lose their 4 neighbours are distributed over this 200τ interval. In Fig. 2 we have plotted the distribution of times at which particles first lost their 4th nearest neighbour. We clearly see that the time at which this condition for irreversibility is first met is strongly weighted to longer times.

Equipped with our measure of irreversibility, we can determine how the *irreversible reorganisation* (IR) is distributed as it originates from a given initial configuration. We record, over an ensemble of 100 iso-configurational runs, the number of runs in which each particle meets the irreversibility criterion within a time interval of 200τ (this corresponds to the time at which the peak of the non-Gaussian parameter [27] occurs at this temperature and represents a time scale of about 2000 collision events). Maps of the log of this probability distribution for six configurations at $T = 0.4$ are shown in Fig. 3. The irreversible reorganization mapped in Fig. 3 are elementary components of the slow structural relaxation characterised by a time $\tau_\alpha$ (= 703τ for the simulations reported here [24]).

**4.2 The Spatial Distribution of Quasi-Localised Normal Modes**

What aspect of the initial configuration is responsible for the observed spatial distribution of irreversible events? As described in Section 2.1, measures associated with local structure have failed to provide strong correlation between the spatial distributions of structure and dynamics in the 2D mixture. To move beyond these purely local measures, we have determined the normal modes for the local potential energy minimum (the 'inherent structure') associated with a number of initial configurations. We shall refer to these as quenched normal modes.

The localization of each mode is measured by the participation ratio,

$P(\omega) = \left[ N \sum_{i=1}^{N} (\vec{e}_\omega^{\,i} \cdot \vec{e}_\omega^{\,i})^2 \right]^{-1}$. If the mode is completely delocalised so all particles contribute equally then $P(\omega) = 1$. At the other extreme, a mode localized on a single particle has $P(\omega) = 1/N$. For a plane wave, $P(\omega) = 2/3$.

The participation fraction of particle $i$ in eigenmode $\mathbf{e}_\omega$ is given by $p_i = |\mathbf{e}^i_\omega|^2$. In Fig. 4 we have plotted maps of the particle participation fractions summed over the 30 lowest frequency modes for each of our initial configurations. This summation corresponds to determining a low temperature local Debye-Waller factor [7,28]. An examination of the individual mode eigenfunctions indicates that these low frequency modes include both localised and delocalised character [11,12,29].

In Fig. 5 maps of the participation fractions for the lowest frequency modes for a single initial configuration are presented along with the corresponding irreversibility map. Quasi-localized modes are defined here as modes with a participation ratio $P < 0.34$ and are coloured red. Heterogeneities are also observed in some delocalized (blue) modes. In all cases, these heterogeneous features of an eigenvector typically consist of a compact group of several particles that contribute significantly to the mode surrounded by an extended 'apron' of particles with smaller contributions. There is a visually striking correlation between regions of motion in the quasi-localized modes and regions of high probability of irreversible reorganisation, which we shall explore further in Section 4.4.

**4.3 Does Directional Motion have a Structural Origin?**

It is common to refer to dynamic heterogeneity as a purely scalar field corresponding to the spatial distribution of relaxation times. Directionality of motion, however, may also be an ingredient of dynamical heterogeneity. This idea has been incorporated into several theoretical approaches aimed at describing dynamical heterogeneity. The question of precisely what aspects of dynamical heterogeneity are spatially anisotropic is somewhat controversial, and must depend on the time scale of observation. Here, we address two questions. First, does a configuration impart some directionality to particles along with a characteristic relaxation time? Second, does the polarization of the low frequency normal modes correlate with the directionality in the motion, if any exists?

We address the first question by constructing a map of single particle displacements, similar to the usual propensity map, but including the vector nature of the direction of displacement. After averaging over an iso-configurational ensemble of trajectories, we indeed find that, at least for relatively short times, the displacement map carries clear directional information. In Fig. 6(a) we present the spatial distribution of the directionality in the form of a vector map. The length of a vector arrow in the map is proportional to the average directional displacement over $10\tau$ after iso-configurational averaging. The coloured contours correspond to the iso-configurational propensity. We find, on these timescales (which corresponds to the middle of the β-relaxation regime), directionality of motion is indeed encoded in the structure of the initial configuration. Establishing the details of how the directionality decays with time are left to future work.

In Fig. 6(b) we also display one particular low frequency quasi-localized normal mode obtained from the same initial configuration used to construct the displacement map. There is a clear visual correspondence between the direction of the mode polarization and the motion that persists in single particle displacements *even after iso-configurational averaging on time scales corresponding to the β regime*. This rather striking result is consistent with observations made previously in slightly different contexts. For example, Brito and Wyart [15,16] have demonstrated that the directional motion that occurs during an "avalanche" leading to relaxation can occur along a single mode in a two-dimensional hard-sphere mixture near the jamming threshold. Since the polarization of an individual mode is usually anisotropic and directional, so too is the motion of the particles. Similar earlier findings were made by Schroeder *et al.* [10-13] and Ashwin *et al.* [30] who displayed the mobility of particle motion in inherent dynamics as the system transits from one inherent structure to an adjacent one. The crucial distinction between our present finding and the earlier ones is that we have explicitly demonstrated a correlation between the directionality of a *statistically averaged quantity* and an intrinsic, non-statistical property (the mode polarization) of the configuration itself. This result allows us to conclude that even in the ensemble-averaged context, heterogeneous motion in the β regime has a directional component. As we will show below, there is a systematic correlation between this directionality and the low frequency normal modes of the system above that demonstrated graphically in Figs. 6a and b.

Directional particle displacement maps are obtained via iso-configurationally averaging over randomly sampled velocity distributions. As such, the correspondence between mode polarization and iso-configurationally averaged displacement is more complex than simply a connection between a single mode and the motion of the configuration as a whole. Indeed, we find a fair amount of redundancy in the polarization of the eigenvectors associated with the particular modes that correlate strongly with maps of propensity and irreversible reorganization. In Fig. 7 we show 4 different normal mode vectors and we compare them to the directionality of the particle map. Regions in blue are correlated with displacement map vectors with an angle of less than $\pi/2$ while the red regions are anti-correlated with angle greater than $\pi/2$. It should be noted that each eigenvector provides us with lines in space along which particles move with the actual forward or backward direction depending on the (arbitrary) phase of the vibration. To make these correlations clearer, we plot the scalar product of the mode polarization with the displacement map field. Darker colors signify greater overlap, with blue (red) again signifying correlation (anticorrelation) with the respective vector fields. Clearly in each case there are well-defined regions of overlap with a distinct phase.

The more global significance of our finding relates to the nature of the directionality of dynamical heterogeneity above and beyond the connection with normal modes. One of the most enduring images of the nature of heterogeneous dynamics in supercooled liquids is that of string-like motion. Evidence is accumulating that this type of directionality is a property of the β regime and not the α regime. Doliwa and Heuer [31] devised a multi-point correlation function for the purposes of detecting particle directionality in a hard-sphere system. They demonstrated that the peak of this function occurred in the β regime, while the motion was predicted to be isotropic in the α regime. Appignanesi *et al.* [32] demonstrated that cooperative clusters in the β regime were compact. Recent experimental work by Candelier *et al.* [33] followed the sequence of motion of grains in dense granular media. Their work suggested that relatively short time motion occurs along normal modes and is directional, while longer time relaxation cascades through the system in a more isotropic manner. Interestingly, inhomogeneous mode-coupling theory [34] is consistent with a change in the geometry of dynamic heterogeneity from fractal to compact as time progresses. In this theory, the change in geometry is reflected in a change in the scaling form associated with multi-point scattering functions. Recent computer simulations quantitatively support the temporal scaling forms predicted by that theory [35]. While in some sense mode-coupling theory is a theory of marginally stable vibrational modes, the intriguing connection suggested by the consistencies noted above certainly deserve future scrutiny.

## 4.4 The Spatial Correlation Between Irreversible Relaxation and Quasi-Localised Modes

To establish the connection between the mode structure and the subsequent irreversible reorganization, we have plotted in Fig. 8 the positions (white circles) of those particles with probability $\geq 0.01$ of meeting our IR condition during the entire $200\tau$ interval in the iso-configurational ensemble on top of the maps of the participation fractions for the low frequency modes at time $t = 0$. Note that the majority contribution to the IR map comes from particles that lost their fourth neighbour late in the trajectory. While these results do not address the question of *when* or even *if* a given soft local mode will become involved in IR, they do strongly support a picture in which the irreversible reorganisation of a configuration originates from these modes.

Two points are worth emphasising. The mode participation fractions, whose spatial distributions are mapped in Figs. 4 and 5, are properties of the *static* initial configurations. Our demonstration of a strong correlation between the mode maps and the irreversible reorganization maps constitutes a significant success in understanding how structure determines relaxation in an amorphous material. Indeed, as Fig. 8 illustrates, one may provide semi-quantitative prediction of IR domains as they emerge at relatively long times from the initial configuration alone. The second point is that, since we have used *quenched* modes, we have only used information about the *bottom* of the local potential minima. While it is quite possible that the time scale required for a reorganisation event will depend on the energy barriers associated with the transition, our results indicate that the spatial structure of such events is largely determined by the distribution of soft quasi-localised modes in the initial configuration, at least in the range of densities and temperatures that can be accessed via computer simulation.

## 4.5 The Evolution of the Spatial Distribution of Quasi-Localised Modes

Given our conclusion that relaxation correlates with soft quasi-localised modes, it follows that our capacity to predict the subsequent spatial distribution of the irreversible relaxation depends on how persistent the mode distribution is in a configuration. After all, should the mode maps evolve rapidly then the structural information in a given configuration would quickly become irrelevant. Krämer et al [36] have discussed the problem of the time evolution of instantaneous normal modes as a means of treating anharmonic effects. The fact that we observe strong spatial correlations between the initial modes and relaxation some $200\tau$ later indicates that the spatial structure of the modes does generally persist over such times. This is remarkable given that small variations in the quenched modes, indicative of a change in the local minimum (or inherent structure), occur over $\sim 1\tau$ intervals.

In Fig. 9 we plot the contour plots of the participation fraction summed over the 30 lowest frequency modes of the quenched configurations generated every $1\tau$ along a $10\tau$ trajectory. We remind the reader that $1\tau$ corresponds to $\sim 10$ collision times. We observe that the participation fraction maps undergo clear changes, indicative of a change in the local quenched potential minimum (or 'inherent structure'), while the overall spatial distribution of local low frequency modes changes little.

We do, however, see examples where the mode structure is not so stable. In Figs. 10a and 10b we compare the mode participation map for the initial configuration with the map of the maximum participation fraction observed per particle over five $10\tau$ runs starting from the configuration in Fig. 10a. The difference in spatial structure between these maps is a measure of the degree of variability of the mode structure. In Figs. 10c and 10d we overlay the particles exhibiting IR within $200\tau$ over the maps of Figs. 10a and 10b, respectively. While the mode structure of the initial configuration does not provide a quantitative predictor of the spatial distribution of IR (see Fig. 10c), the cumulative mode structure sampled over the multiple short runs does (see Fig. 10d). This result demonstrates that even when the soft mode structure is not stable, the irreversible relaxation still originates with these soft modes, only now this IR is not well predicted by any single configuration. It appears that configurations such as that analysed in Fig. 10 represent those caught in transit between configurations with more stable mode structure.

Unlike the configuration shown in Fig. 10, many configurations show little significant difference between the low frequency participation map generated from the initial configuration and the map obtained by recording the maximum value of the participation fraction of each particle achieved over five $10\tau$ runs. An example is the configuration shown in Fig. 9. We conclude that the spatial structure of the quenched soft modes can often persist over many changes in the inherent structure. Preliminary results indicate that this persistence is also found in 3D mixtures (including temperatures below the empirical mode coupling temperature) (see Appendix).

## 4.6 The Overlap of Low Frequency Quenched Modes and Imaginary Instantaneous Modes

One aspect that can be questioned regarding the relationship of normal modes to dynamics as presented in this work is the reliance on quenched normal modes. Since true dynamical trajectories do not access the configurations located by the quench procedure, the relationship between these modes and real dynamics is not entirely clear. In particular, at relatively high temperatures, the utility of inherent structures is questionable. In several works, a connection between imaginary frequency instantaneous normal modes and glassy behaviour has been made. Most pertinent to our study is the recent work of Coslovich and Pastore [37]. These authors found a relationship between localized saddle modes and spatial regions of high propensity.

If we choose to focus on the putative relationship between instantaneous normal modes and heterogeneous dynamics, we are immediately confronted with a problem. The mode map created from the quenched modes may be viewed as an effective harmonic Debye-Waller map (see Section 2.1). For the instantaneous normal modes, one might naively view the real frequency modes as the proxy for the quenched modes that form this map. However, we have found that there is essentially no correlation between the real instantaneous modes and the irreversibly relaxing particles. On the other hand, there does exist interesting correlations between the imaginary frequency modes and the various iso-configurational maps. If we crudely sum the imaginary frequency mode participation numbers as we have done for the quenched modes, we find maps that do correlate with the propensity, Debye-Waller, and irreversibly relaxing particle maps. In Fig. 11 we show the Debye-Waller, quenched and imaginary instantaneous normal mode maps, respectively. There is a clear visual correspondence, which illustrates the fact that while the overlap between the imaginary instantaneous mode and the Debye-Waller maps is weaker than the overlap between the quenched mode and the Debye-Waller maps, a clear correlation does exist. The weaker correlation for the instantaneous modes appears to arise from a greater degree of localization in these modes. We quantify both of these relationships in Table 1.

Presumably, the relation between the real quenched modes and the imaginary instantaneous modes arises because the softest directions about a local minimum are continuously connected imaginary modes away from stable points on the landscape. In this regard, our results are consistent with the earlier work of Coslovich and Pastore [37]. It should be noted, however, that the relationship between generic imaginary modes and saddle modes is not clear. It is interesting that in the low temperature regime studied here, the quenched normal mode approach provides a more accurate reconstruction of the various iso-configurational maps. It would be interesting to investigate the possibility that the instantaneous normal mode approach might provide a better picture of iso-configurational properties at the higher temperatures where dynamically heterogeneous behaviour first manifests.

**5. Conclusions**

In this paper we have presented two important results relating to slow relaxation in a model supercooled liquid. The first is that the irreversible reorganisation originates at the sites of low frequency quasi-localised quenched modes. The second is that these modes typically persist for time scales significantly longer than the lifetime of a given inherent structure. These results show that the spatial location and extent of IR regions at relatively long times may be reasonably predicted by a simple, static property of the static initial condition.

What exactly is the relationship between the local soft modes and the eventual reorganization? The approach to the glass transition is characterized by the increasing degree to which structure determines the collective dynamics[38]. Soft modes represent a direct and, currently, unique signature of that aspect of the structure that determines the position and nature of the localised processes by which the liquid moves between inherent structures. As pointed out by Ashton and Garrahan, this fact alone does not mean that the soft-modes are directly involved in the relaxation process [39]. These authors argue that the soft-modes may simply mark the location of defects whose kinetic mobility is determined by other factors. On the other hand, Brito and Wyart have shown rather convincing visual evidence that, at the very least, the direction of motion during relaxation events strongly correlates with a small number of normal mode polarizations. This is consistent with the robust (after iso-configurational averaging) behaviour we observe here demonstrating strong correlations between motion on the $\beta$ scale and mode directionality. These results suggest that the soft-modes themselves are indeed the origin of relaxation phenomena and not merely passive bystanders. However, more work is required to completely determine the degree of *active participation* of soft-modes in irreversible relaxation.

As we have emphasised, the persistence of the spatial heterogeneity of the sum of the low frequency modes in the face of rapid fluctuations in the identity of individual modes is an essential feature by which a specific configuration influences relaxation dynamics over a substantial time interval. Recently, Brito and Wyart [40] have reported a more stringent persistence of individual modes in a hard disk mixture close to jamming. They conclude that the each mode remains unchanged until the point where a single extensive mode goes unstable and an 'avalanche' ensues. This picture is not entirely consistant with the short time scale fluctuations we observe in our thermal system. It is not clear if this persistence marks a fundamental distinction between the landscapes of hard-sphere and thermal systems, or simply a distinction between the time-averaged free-energy modes used by Brito and Wyart and the quenched modes used here.

The results of this paper suggest that the quasi-localised modes can provide a unification of many of the major themes in current research on the glass transition. Can the growth in the four point susceptibility $\chi_4$ near the glass transition [41] and the jamming transition in granular material [42], along with the growth in the related kinetic correlation length, be directly connected to an equilibrium correlation function of the soft mode distribution[43]? Does the persistence of the mode structure in real space reflect the transient confinement of the system trajectory within a 'metabasin' in configuration space [8,44]?

At the most fundamental level, these quasi-localised modes represent the strongest link yet established between structure and dynamic heterogeneity and, hence, an exciting route forward to establish how molecular properties influence relaxation in the supercooled liquid.

**Acknowledgements**

We would like to thank L. Berthier, G. Biroli, J. P. Bouchaud, A. Heuer, and C. O'Hern for useful discussions. HP and DRR would like to thank P. Verrocchio for providing the equilibrated 3D configurations and the NSF for financial support. AW and PH acknowledge the support of the Australian Research Council.

APPENDIX

**On the Similarity of Structure-Dynamics Correlations in 2D and 3D**

There is a striking correlation between the regions of motion in the quasi-localized modes ($P_\omega \leq 0.34$) and the regions of high Debye-Waller factors [7] in both 2D and 3D. To quantify this, we look at particles with the top 10% highest values of the local Debye-Waller factor compared to the top 10% with the highest low-frequency motion. The low-frequency motion of each particle is obtained by summing its participation fraction, $p_i$, for all low frequency modes. (As the number of low frequency plane wave modes increases with system size, some upper limit in participation ratio may need to be applied for larger systems.) Summing between 30-80 of the lowest frequency modes gives qualitatively similar results. The overlap of the regions of large short-time displacement large normal mode displacement are computed by first eliminating particles from both maps that are not in a cluster of at least $n^* = 3$ to eliminate noise and focus on regions of collective motion. The percentage of particles in clusters of the top 10% for both the local Debye-Waller factor and the summed 30 lowest frequency modes for 10 different configurations are plotted in Fig. A1, along with the overlap of clusters of a random sample of 10% of the particles for comparison.

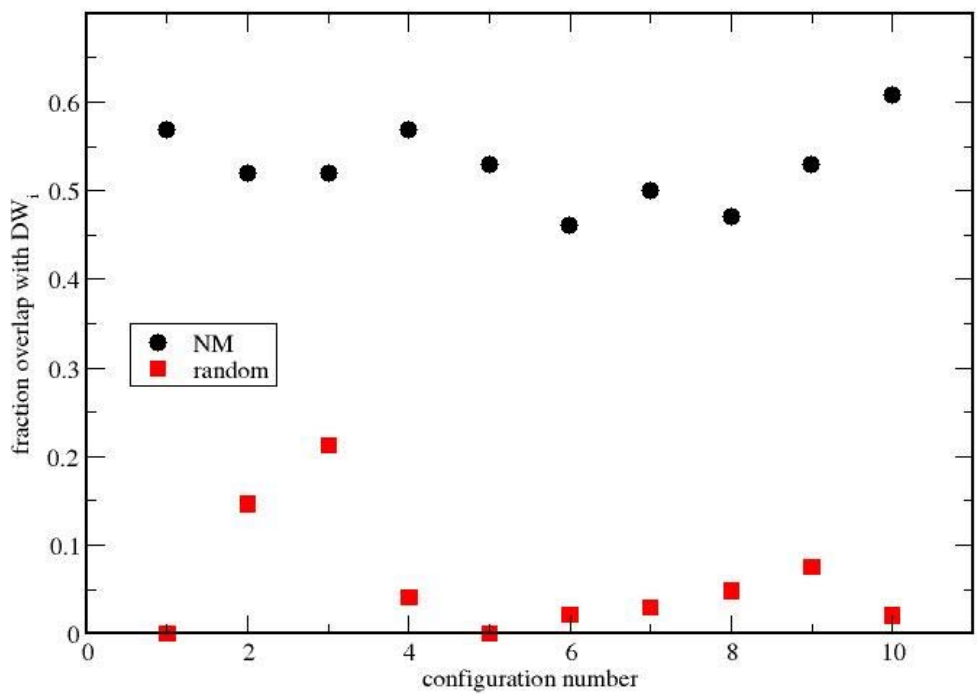

**Figure A1**.  For the 2D system, the overlap of the 10% of particles with the largest Debye-Waller factors with (black dots) the top 10% of clustered particles with the largest amplitudes in the sum of the lowest 30 normal modes and with (red squares) a random selection of 10% of the particles, all maps have had clusters of less than $n^*=3$ eliminated.

To insure that this correlation holds in 3D, we performed a similar analysis on a 3D binary soft-sphere system. These particles interact with a pairwise repulsion,

$$u_{ab}(r) = \varepsilon \left[ \frac{\sigma_{ab}}{r} \right]^{12} \hspace{3cm} \text{(A1)}$$

with $\sigma_{12} = 1.1 \times \sigma_{11}$, $\sigma_{22} = 1.2 \times \sigma_{11}$ and all units reduced such that $\sigma_{11} = 1.0$, $\varepsilon = 1.0$, $m = 1.0$ and the time unit $\tau = \sigma_{11}(m/\varepsilon)^{1/2}$. Equilibrated configurations were obtained from P. Verrocchio [45] for temperature down to $0.92T_{MC}$, where the mode-coupling temperature [46] is $T_{MC} = 0.2084 \ k_B/\varepsilon$. A microcanonical molecular dynamics simulation was carried out for $N = 1024$ particles (50% each small and large)) at a density $\rho = 0.7421 \ \sigma_{11}^{-3}$ and temperature $T = 0.2084 \ k_B/\varepsilon$ in a cube box with periodic boundary conditions. The $\alpha$ relaxation time was found to be $\tau_\alpha = 832 \ \tau$ and the non-Gaussian parameter peak at $200\tau$. The local Debye-Waller factor for each particle was calculated as the variance of 100 runs of the iso-configurational distribution of each particle's displacement over a time interval of $10\tau$. The same normal-mode analysis was performed on this system as on the 2D system, but the dimensionality of the eigenspace is 50% larger than for the 2D system. In 3D, analyzing the overlap of the clusters of the 10% of particles with the largest local Debye-Waller factor and those with the highest summed amplitude from the low-frequency modes show that anywhere between 45-150 modes gives a good prediction of the short-time dynamics. Clusters in 3D are defined with a minimum size of $n^* = 5$. Fig. A2 shows the overlap between the most mobile particles at $10\tau$ and the summed amplitudes of the 60 lowest frequency modes.

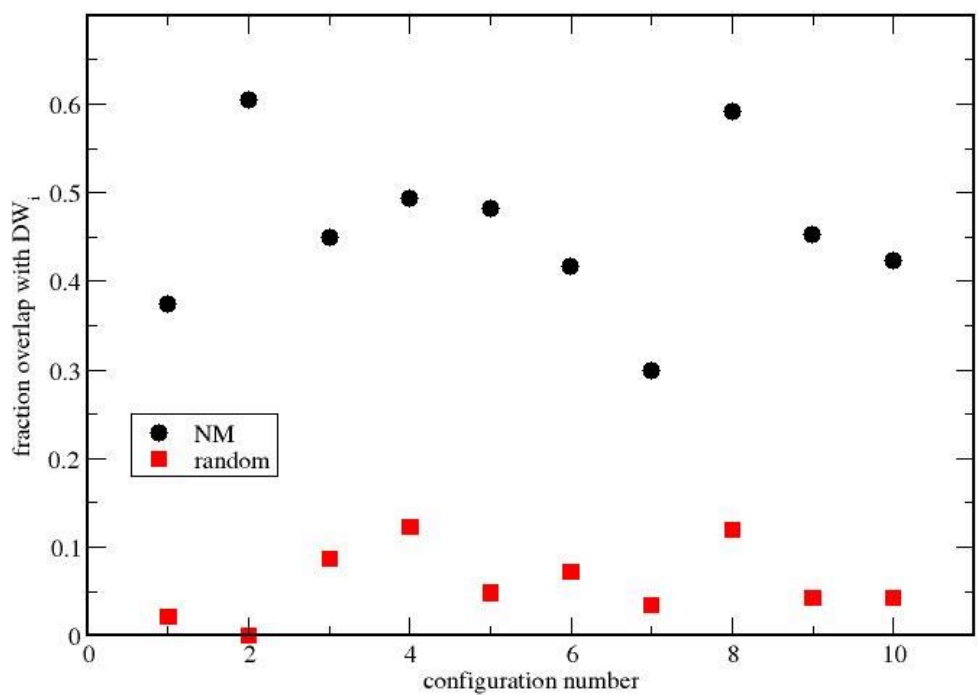

**Figure A2.** Similar to Figure S6 for the 3D system, in this dimension the minimum cluster size is $n^*=5$.

**Figure Captions**

**Figure 1** The probability distributions for the minimum number of 'lost' initial neighbours $n$ after having first lost $x$ neighbours. Note that after losing $x = 4$ neighbours, the probability of recovering all of the initial neighbours (i.e. having $n = 0$) is essentially zero. The distributions show peaks at $n = x\text{-}2$ suggesting that particles typically recover 2 of their initial neighbours. These calculations where carried out over a time interval of $1000\tau$ at $T = 0.4$.

**Figure 2.** Distribution of times at which particles first lose their 4th nearest neighbour (data averaged over isoconfigurational ensembles of 100 runs for a total of 10 initial configurations of $N = 1024$ particles at $T = 0.4$).

**Figure 3.** Contour plots of the probability ($\log_{10}$) of a particle losing 4 original neighbours, the criterion for irreversible reorganization, over 100 iso-configurational runs for 6 different initial configurations.

**Figure 4.** Contour plots of the participation fraction summed over the 30 lowest frequency modes for the quenched initial configurations of the same 6 configurations used in Fig. 3.

**Figure 5**. Maps of the participation fractions for the 11 lowest frequency normal modes of the local potential energy minimum associated with a single initial configuration. Modes with a participation ratio $P < 0.34$ are coloured red while the more delocalized modes are coloured blue. The intensity of the colour increases with the magnitude of the squared amplitude. The corresponding irreversibility map for this configuration is provided for comparison.

**Figure 6.** (Left panel)) A map of the iso-configurational average of the displacement vector (not the square of the displacement as in the propensity) plotted on top of the respective $10\tau$ propensity map. (Right panel) A map of the polarization vectors of a single low frequency normal mode for the initial configuration used in the left panel.

**Figure 7.** A comparison of the particle displacements averaged over the iso-configurational ensemble and the displacements associated with the 4 lowest frequency normal modes. The first column labelled 'Displacements' contains the same displacement map repeated for ease of comparison. The vectors are proportional to the iso-configurational average of the particle displacement over $10\tau$. The middle column labelled "Mode Displacements" shows the 4 lowest frequency normal modes of the initial configuration. The vectors are proportional to the mode polarization vector on each particle. Those vectors coloured blue are correlated with displacement map vectors with an angle of less than $\pi/2$ while the red regions are anti-correlated with angle greater than $\pi/2$. It should be noted that each eigenvector provides us with lines in space along which particle moves with the actual forward or backward direction depending on the (arbitrary)

phase of the vibration. The third column labelled "Comparison" shows plots of the scalar product of the mode polarization vector on each particle with the average displacement vector. Darker colors signify greater magnitude of the scalar product, with blue (red) signifying positive (negative) signs of the scalar product.

**Figure 8.** Contour plots of the low frequency mode participation (as in Fig. 3), overlaid with the location of particles (white circles) whose iso-configurational probability of losing 4 initial nearest neighbours within $200\tau$ is greater than or equal to 0.01.

**Figure 9.** Plots of the participation fraction in the 30 lowest frequency normal modes for quenched configurations taken every $1\tau$ along a $10\ \tau$ trajectory from the initial configuration. The colour code is the same as previous figures. While there are clearly variations occurring in the distribution of modes (and hence in the identity of the quenched minimum or inherent structure) over $1\tau$, substantial elements of the mode distribution persist. The top left map is the irreversible reorganization map for the initial configuration, included for comparison.

**Figure 10.** a) Contour plot of the participation fraction summed over the 30 lowest frequency modes for a quenched configuration. b) Contour plot of the maximum value of the participation fraction observed over five $10\tau$ runs starting from the configuration in Fig. 10a. c) Particles whose iso-configurational probability of losing 4 initial nearest neighbours within $200\tau$ is greater than or equal to 0.01 (white circles) overlaid on the participation fraction map for the initial configuration. d) As in c) except that the overlay is over the map of the maximum participation fraction shown in b).

**Figure 11.** A comparison of the sum of the participation fractions of the normal modes with imaginary frequency with the sum of participation fraction of the same number of quenched normal modes, along with the map of the local Debye-Waller factor and a map with vectors representing the motion of each particle from the instantaneous configuration (IC) to the inherent structure (IS).

| QNM (%) | INM (%) | overlap (%) |
|---------|---------|-------------|
| 78 | 8 | 92 |
| 41 | 14 | 59 |
| 43 | 15 | 64 |
| 27 | 11 | 50 |
| 49 | 13 | 75 |
| 60 | 7 | 89 |
| 75 | 9 | 89 |
| 73 | 14 | 91 |
| 40 | 14 | 62 |
| 72 | 7 | 90 |

**Table 1.** Comparison of the degree of similarity in the sum of low-frequency quenched normal modes and imaginary frequency instantaneous normal mode maps for 10 configurations. The left two columns give the percentage of particles with a value of the amplitude of the sum of the mode polarization vectors on each particle greater than 25% of the minimum value (the "active portions"). The overlap column lists the percentages of those particles that are presented in both the left and middle columns.

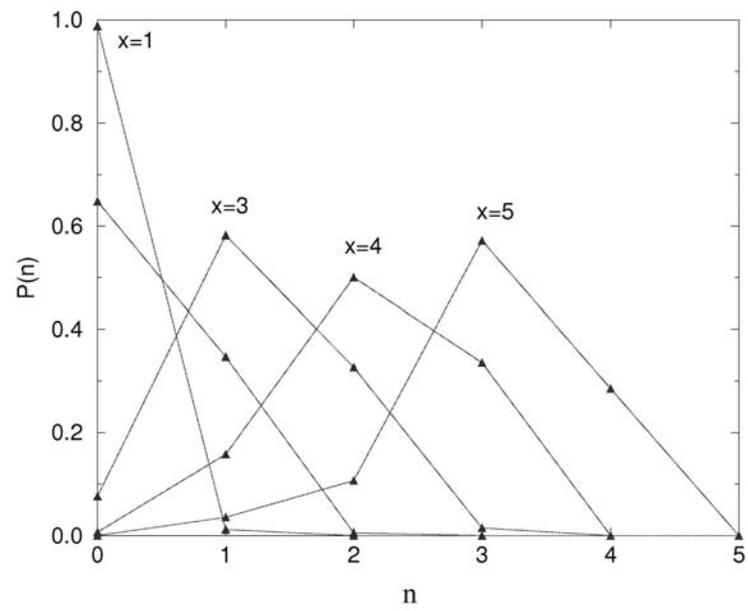

Figure 1

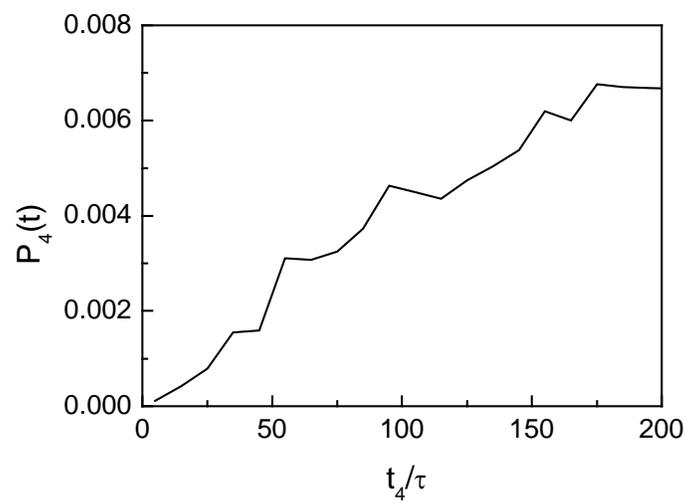

Figure 2

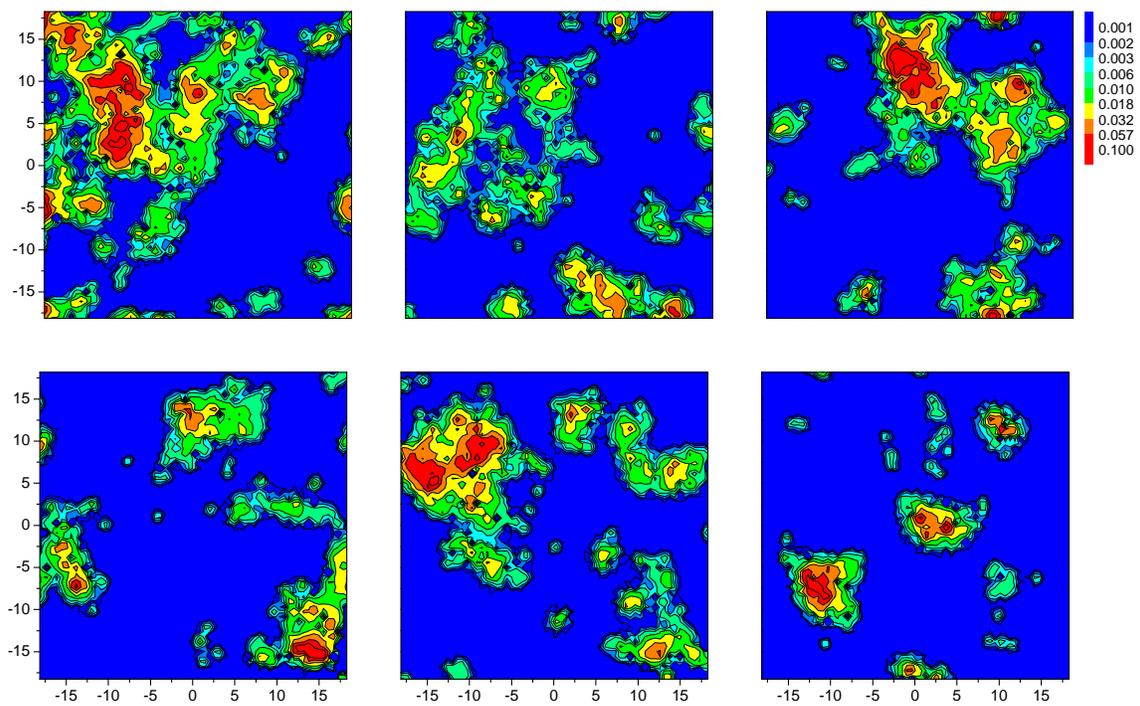

Figure 3

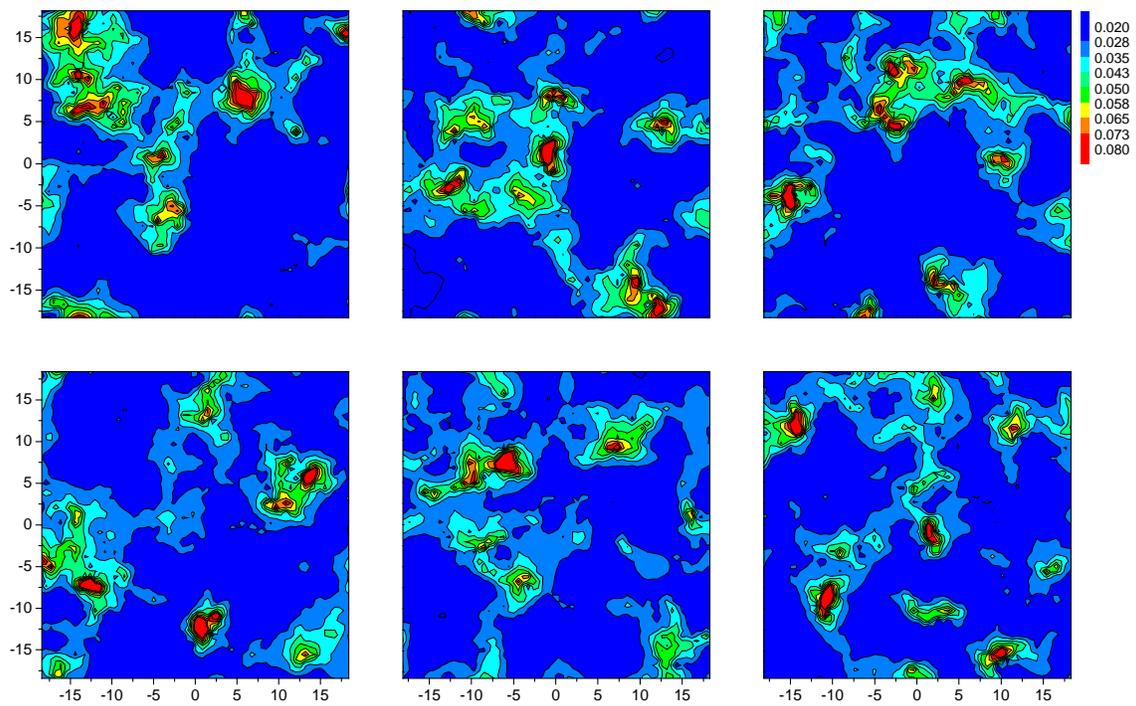

Figure 4

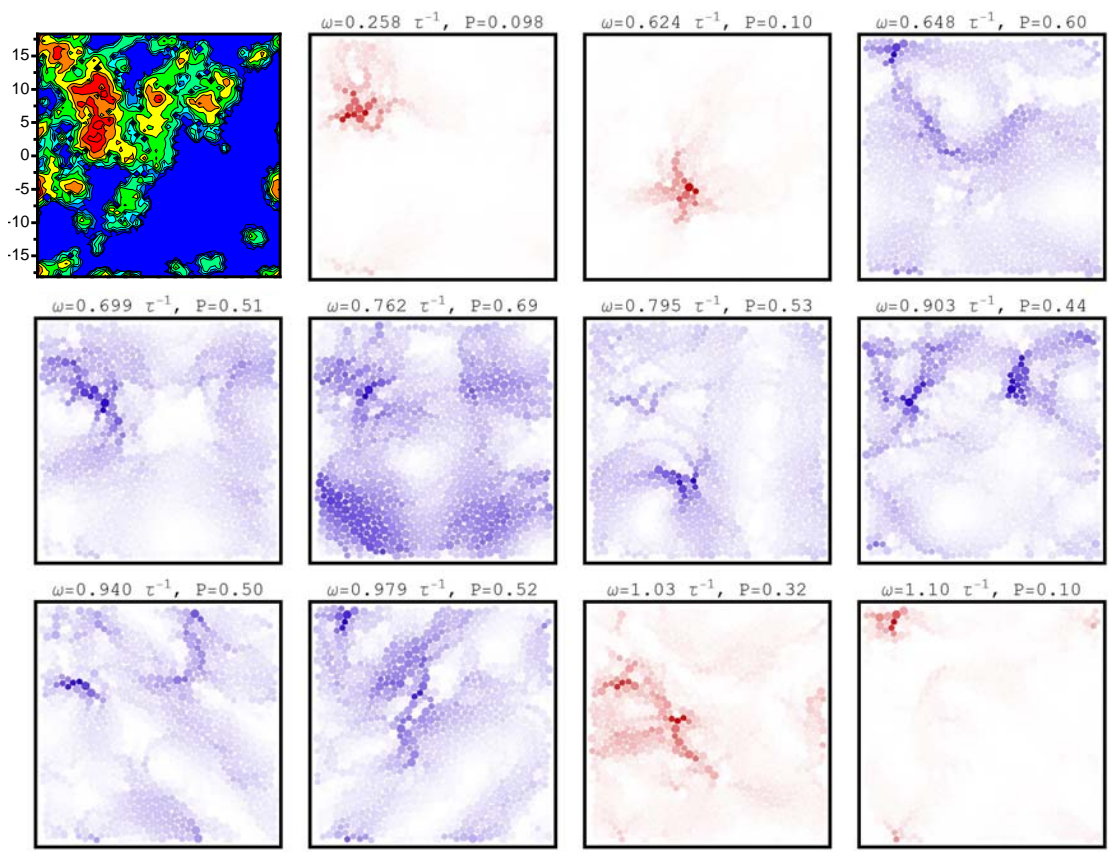

Figure 5

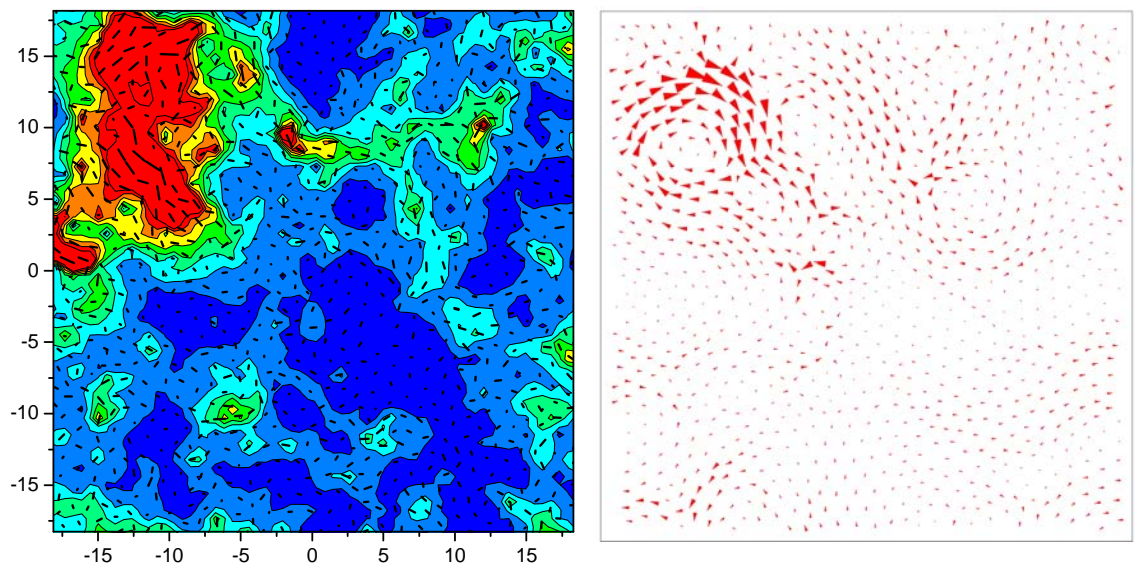

Figure 6

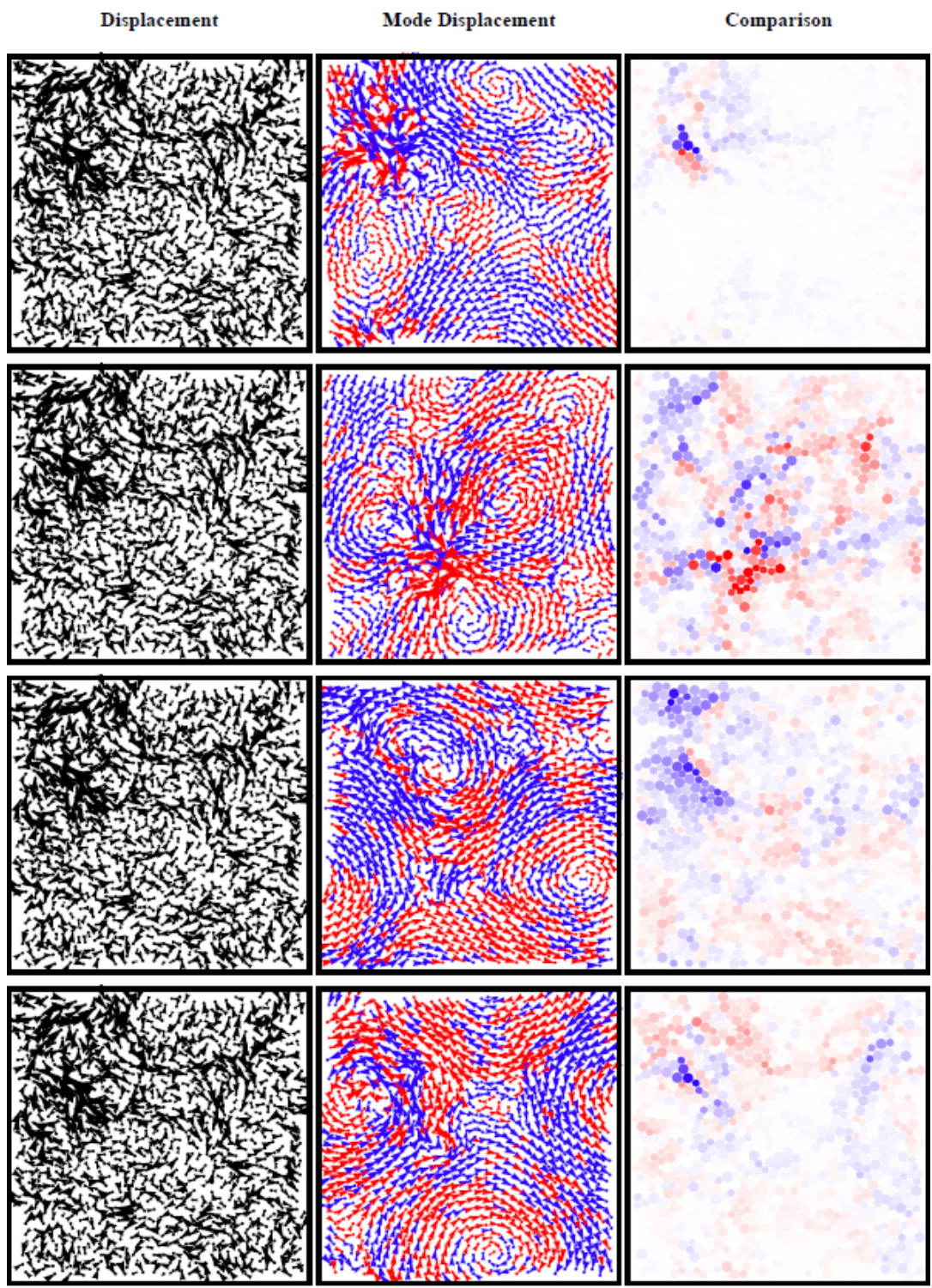

Figure 7

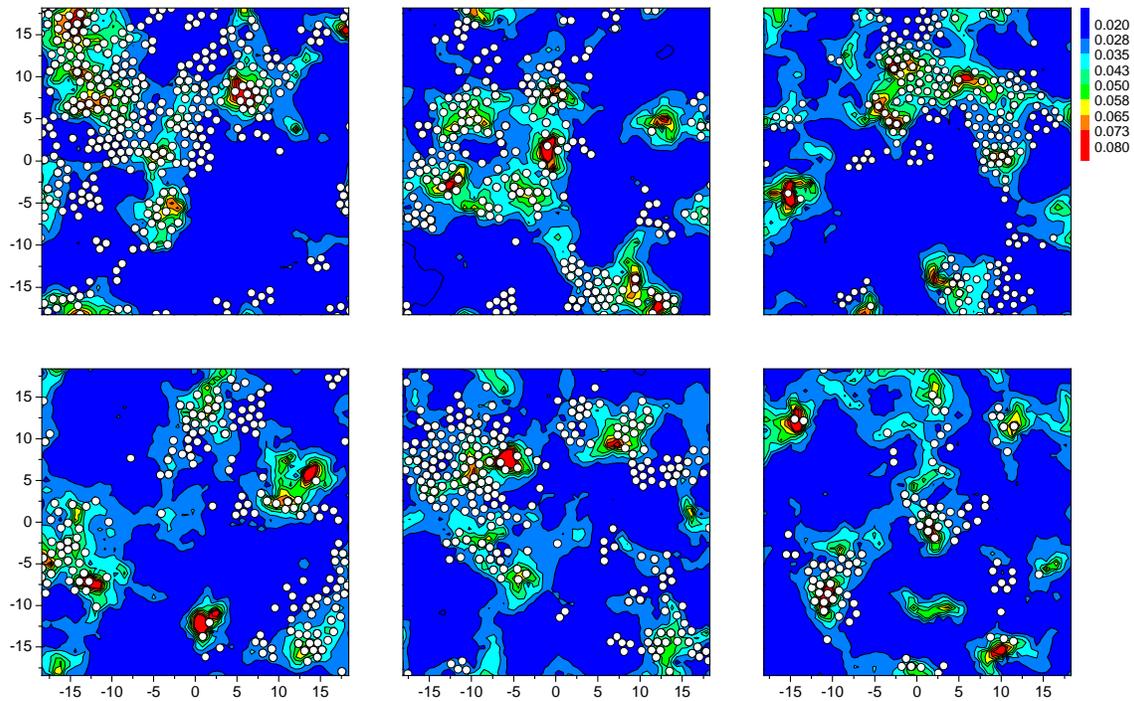

Figure 8

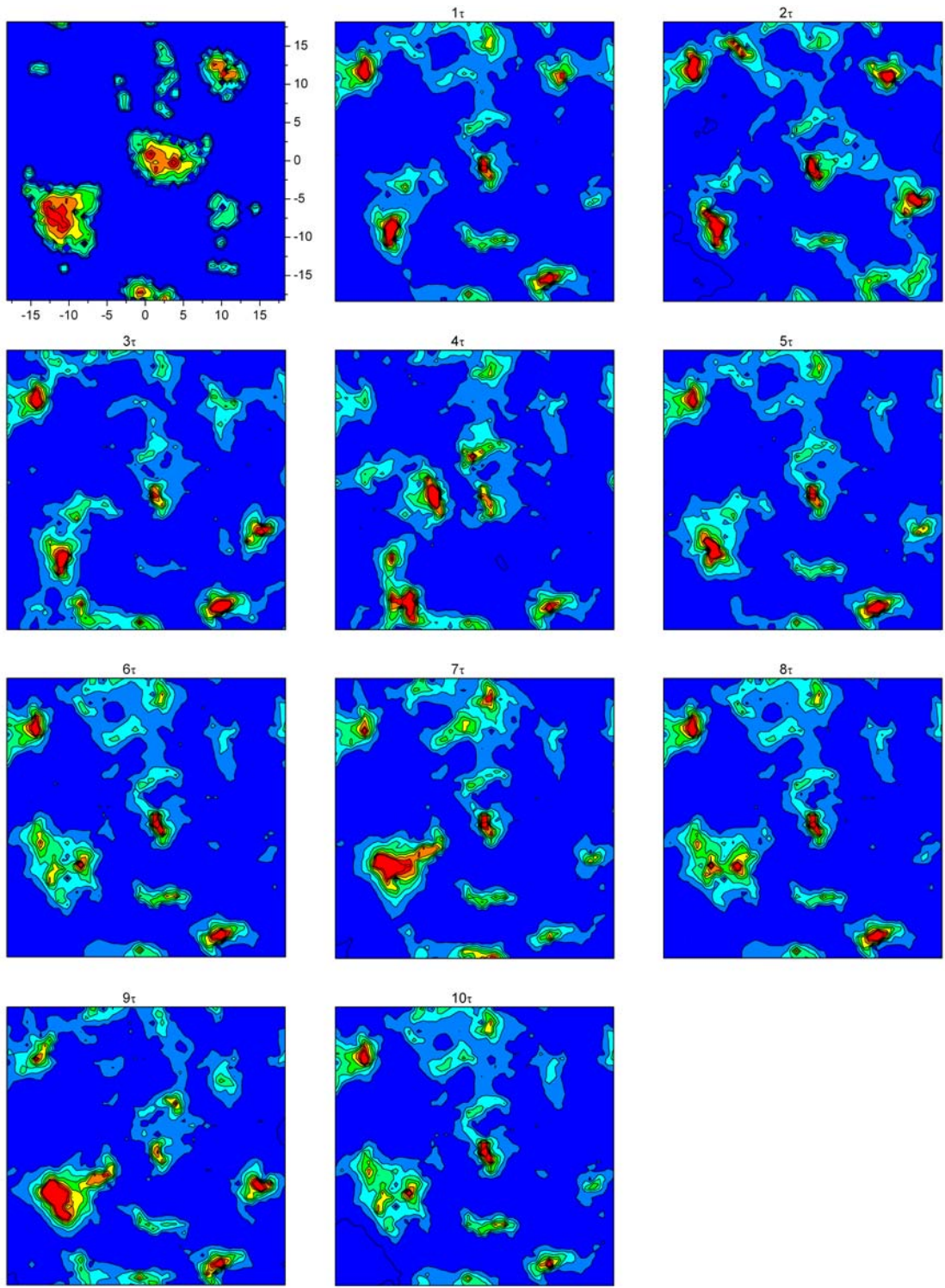

Figure 9

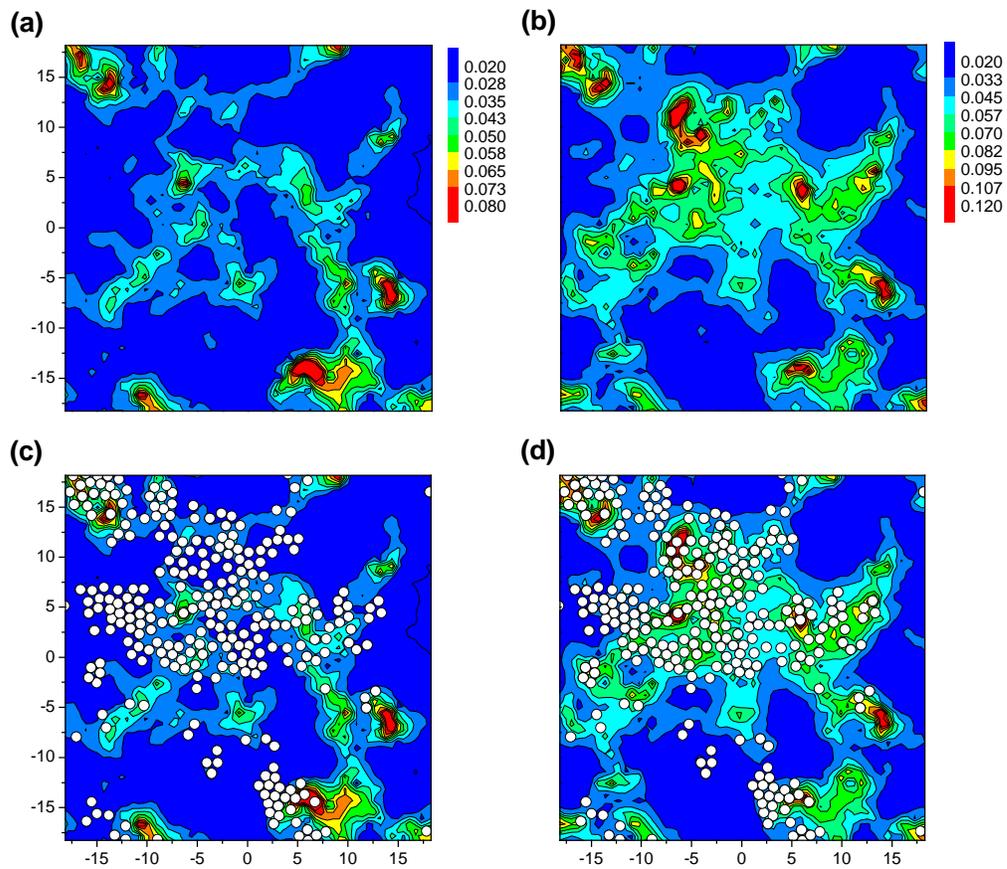

Figure 10

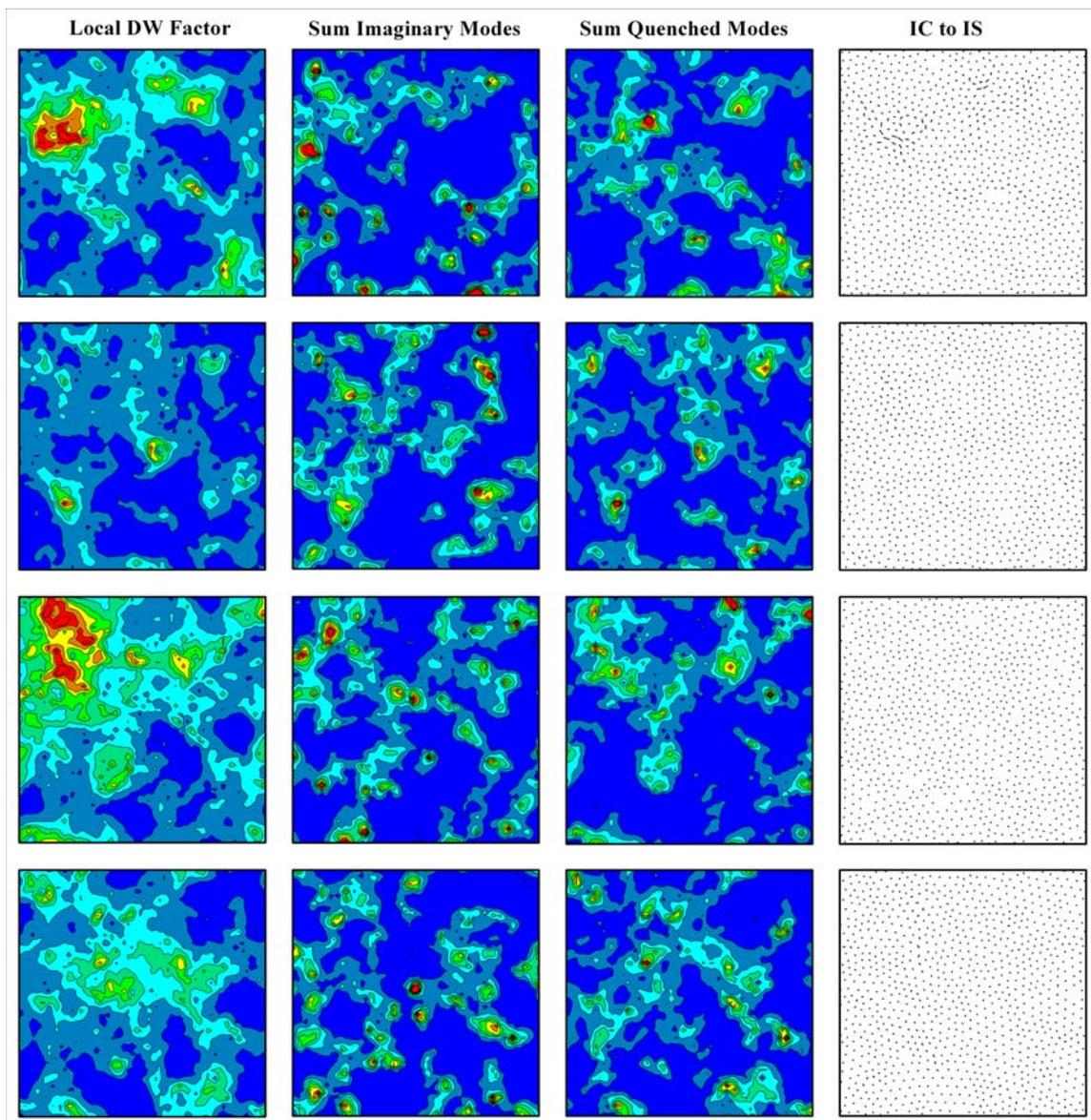

Figure 11